
\magnification=1200
\hsize=31pc
\vsize=55 truepc
\hfuzz=2pt
\vfuzz=4pt
\pretolerance=5000
\tolerance=5000
\parskip=0pt plus 1pt
\parindent=16pt
\font\fourteenrm=cmr10 scaled \magstep2
\font\fourteeni=cmmi10 scaled \magstep2
\font\fourteenbf=cmbx10 scaled \magstep2
\font\fourteenit=cmti10 scaled \magstep2
\font\fourteensy=cmsy10 scaled \magstep2
\font\large=cmbx10 scaled \magstep1

\font\eightrm=cmr8
\font\eighti=cmmi8
\font\eightbf=cmbx8
\font\eightit=cmti8

\font\eightsy=cmsy8
\font\sixrm=cmr6
\font\sixi=cmmi6
\font\sixsy=cmsy6

\def\tenpoint{\def\rm{\fam0\tenrm}%
  \textfont0=\tenrm \scriptfont0=\sevenrm
                      \scriptscriptfont0=\fiverm
  \textfont1=\teni  \scriptfont1=\seveni
                      \scriptscriptfont1=\fivei
  \textfont2=\tensy \scriptfont2=\sevensy
                      \scriptscriptfont2=\fivesy
  \textfont3=\tenex   \scriptfont3=\tenex
                      \scriptscriptfont3=\tenex
  \textfont\itfam=\tenit  \def\it{\fam\itfam\tenit}%
  \textfont\slfam=\tensl  \def\sl{\fam\slfam\tensl}%
  \textfont\bffam=\tenbf  \scriptfont\bffam=\sevenbf
                            \scriptscriptfont\bffam=\fivebf
                            \def\bf{\fam\bffam\tenbf}%
  \normalbaselineskip=20 truept
  \setbox\strutbox=\hbox{\vrule height14pt depth6pt
width0pt}%
  \let\sc=\eightrm \normalbaselines\rm}
\def\eightpoint{\def\rm{\fam0\eightrm}%
  \textfont0=\eightrm \scriptfont0=\sixrm
                      \scriptscriptfont0=\fiverm
  \textfont1=\eighti  \scriptfont1=\sixi
                      \scriptscriptfont1=\fivei
  \textfont2=\eightsy \scriptfont2=\sixsy
                      \scriptscriptfont2=\fivesy
  \textfont3=\tenex   \scriptfont3=\tenex
                      \scriptscriptfont3=\tenex
  \textfont\itfam=\eightit  \def\it{\fam\itfam\eightit}%
  \textfont\bffam=\eightbf  \def\bf{\fam\bffam\eightbf}%
  \normalbaselineskip=16 truept
  \setbox\strutbox=\hbox{\vrule height11pt depth5pt width0pt}}
\def\fourteenpoint{\def\rm{\fam0\fourteenrm}%
  \textfont0=\fourteenrm \scriptfont0=\tenrm
                      \scriptscriptfont0=\eightrm
  \textfont1=\fourteeni  \scriptfont1=\teni
                      \scriptscriptfont1=\eighti
  \textfont2=\fourteensy \scriptfont2=\tensy
                      \scriptscriptfont2=\eightsy
  \textfont3=\tenex   \scriptfont3=\tenex
                      \scriptscriptfont3=\tenex
  \textfont\itfam=\fourteenit  \def\it{\fam\itfam\fourteenit}%
  \textfont\bffam=\fourteenbf  \scriptfont\bffam=\tenbf
                             \scriptscriptfont\bffam=\eightbf
                             \def\bf{\fam\bffam\fourteenbf}%
  \normalbaselineskip=24 truept
  \setbox\strutbox=\hbox{\vrule height17pt depth7pt width0pt}%
  \let\sc=\tenrm \normalbaselines\rm}

\def\today{\number\day\ \ifcase\month\or
  January\or February\or March\or April\or May\or June\or
  July\or August\or September\or October\or November\or
December\fi
  \space \number\year}

\newcount\secno      
\newcount\subno      
\newcount\subsubno   
\newcount\appno      
\newcount\tableno    
\newcount\figureno   

\normalbaselineskip=20 truept
\baselineskip=20 truept
\def\title#1
   {\vglue1truein
   {\baselineskip=24 truept
    \pretolerance=10000
    \raggedright
    \noindent \fourteenpoint\bf #1\par}
    \vskip1truein minus36pt}
\def\author#1
  {{\pretolerance=10000
    \raggedright
    \noindent {\large #1}\par}}
\def\address#1
   {\bigskip
    \noindent \rm #1\par}
\def\shorttitle#1
   {\vfill
    \noindent \rm Short title: {\sl #1}\par
    \medskip}

\def\pacs#1
   {\noindent \rm PACS number(s): #1\par
    \medskip}
\def\jnl#1
   {\noindent \rm Submitted to: {\sl #1}\par
    \medskip}
\def\date
   {\noindent Date: \today\par
    \medskip}
\def\beginabstract
   {\vfill\eject
    \noindent {\bf Abstract. }\rm}
\def\keyword#1
   {\bigskip
    \noindent {\bf Keyword abstract: }\rm#1}
\def\endabstract
   {\par
    \vfill\eject}

\def\entry#1#2#3
   {\noindent
    \hangindent=20pt
    \hangafter=1
    \hbox to20pt{#1 \hss}#2\hfill #3\par}
\def\subentry#1#2#3
   {\noindent
    \hangindent=40pt
    \hangafter=1
    \hskip20pt\hbox to20pt{#1 \hss}#2\hfill #3\par}
\def\section#1
   {\vskip0pt plus.1\vsize\penalty-250
    \vskip0pt plus-.1\vsize\vskip24pt plus12pt minus6pt
    \subno=0 \subsubno=0
    \global\advance\secno by 1
    \noindent {\bf \the\secno. #1\par}
    \bigskip
    \noindent}
\def\subsection#1
   {\vskip-\lastskip
    \vskip24pt plus12pt minus6pt
    \bigbreak
    \global\advance\subno by 1
    \subsubno=0
    \noindent {\sl \the\secno.\the\subno. #1\par}
    \nobreak
    \medskip
    \noindent}

\def\appendix#1
   {\vskip0pt plus.1\vsize\penalty-250
    \vskip0pt plus-.1\vsize\vskip24pt plus12pt minus6pt
    \subno=0
    \global\advance\appno by 1
    \noindent {\bf Appendix
#1\par}
    \bigskip
    \noindent}

\def\figures
   {\vfill\eject
    \noindent {\bf Figure captions\par}
    \bigskip}
\def\figcaption#1
   {\global\advance\figureno by 1
    \noindent {\bf Figure \the\figureno.} \rm#1\par
    \bigskip}
\def\references
     {\vfill\eject
     {\noindent \bf References\par}
      \parindent=0pt
      \bigskip}
\def\refjl#1#2#3#4
   {\hangindent=16pt
    \hangafter=1
    \rm #1
   {\frenchspacing\sl #2
    \bf #3}
    #4\par}
 \def\numrefjl#1#2#3#4#5
   {\parindent=40pt
    \hang
    \noindent
    \rm {\hbox to 30truept{\hss #1\quad}}#2
   {\frenchspacing\sl #3\/
    \bf #4}
    #5\par\parindent=16pt}
\def\refbk#1#2#3
   {\hangindent=16pt
    \hangafter=1
    \rm #1
   {\frenchspacing\sl #2}
    #3\par}
\def\numrefbk#1#2#3#4
   {\parindent=40pt
    \hang
    \noindent
    \rm {\hbox to 30truept{\hss #1\quad}}#2
   {\frenchspacing\sl #3\/}
    #4\par\parindent=16pt}

\catcode`\@=11
\def\vfootnote#1{\insert\footins\bgroup
    \interlinepenalty=\interfootnotelinepenalty
    \splittopskip=\ht\strutbox 
    \splitmaxdepth=\dp\strutbox \floatingpenalty=20000
    \leftskip=0pt \rightskip=0pt \spaceskip=0pt \xspaceskip=0pt
    \noindent\eightpoint\rm #1\ \ignorespaces\footstrut\futurelet\next\fo@t}
\def\ind{\hbox to 5pc{}}
\def\eq(#1){\hfill\llap{(#1)}}

\def\deqn#1{\displ@y\halign{\hbox to \displaywidth
    {$\@lign\displaystyle##\hfil$}\crcr #1\crcr}}
\def\indeqn#1{\displ@y\halign{\hbox to \displaywidth
    {$\ind\@lign\displaystyle##\hfil$}\crcr #1\crcr}}
\def\indalign#1{\displ@y \tabskip=0pt
  \halign to\displaywidth{\ind$\@lign\displaystyle{##}$\tabskip=0pt
    &$\@lign\displaystyle{{}##}$\hfill\tabskip=\centering
    &\llap{$\@lign##$}\tabskip=0pt\crcr
    #1\crcr}}
\catcode`\@=12
\def\JPA{J. Phys. A: Math. Gen.}

\def\JETP{Sov. Phys. JETP}

\def\PLA{Phys. Lett. A}

\def\PRB{Phys. Rev. B}
\def\PRL{Phys. Rev. Lett.}

\def\PRS{Proc. R. Soc.}

\def\ZPB{Z. Phys. B}

\message{cross referencing macros - BD 1991}

\catcode`@=11
\immediate\newread\xrffile\immediate\openin\xrffile=\jobname.xrf
\ifeof\xrffile
  \message{ no file \jobname.xrf - run again for correct forward references }
\else
  \immediate\closein\xrffile\input\jobname.xrf
\fi
\immediate\newwrite\xrffile\immediate\openout\xrffile=\jobname.xrf

\newcount\t@g
\def\order#1{\expandafter\expandafter\csname newcount\endcsname
   \csname t@g#1\endcsname\csname t@g#1\endcsname=0
   \expandafter\expandafter\csname newcount\endcsname
   \csname t@ghd#1\endcsname\csname t@ghd#1\endcsname=0

   \expandafter\def\csname #1\endcsname##1{\csname next#1\endcsname##1 }

   \expandafter\def\csname next#1\endcsname##1 %
     {\edef\t@g{\csname t@g#1\endcsname}\edef\t@@ghd{\csname
t@ghd#1\endcsname}%
      \ifnum\t@@ghd=\t@ghd\else\global\t@@ghd=\number\t@ghd\global\t@g=0\fi%
     \ifunc@lled{@#1}{##1}\global\advance\t@g by 1%
       {\def\next{##1}\ifx\next\empty%
       \else\writenew{#1}{##1}\expandafter%
       \xdef\csname @#1num##1\endcsname{\t@ghead\number\t@g}\fi}%
       {\t@ghead\number\t@g}%
     \else
       \message{ Warning - duplicate #1 label >> ##1 << }
       \csname @#1num##1\endcsname%
     \fi}%

   \expandafter\def\csname ref#1\endcsname##1{%
     \expandafter\each@rg\csname #1eatspace\endcsname{##1}}

   \expandafter\def\csname #1eatspace\endcsname##1 ##2,%
    {\csname #1cite\endcsname##2 ##1 }

   \expandafter\def\csname #1cite\endcsname##1 ##2##3
     {\ifunc@lled{@#1}{##2##3}%
       {\expandafter\ifx\csname @@#1num##2##3\endcsname\relax%
         \message{ #1 label >>##2##3<< is undefined }>>##2##3<<%
       \else\csname @@#1num##2##3\endcsname##1\fi}%
     \else\csname @#1num##2##3\endcsname##1%
     \fi}}

\def\each@rg#1#2{{\let\thecsname=#1\expandafter\first@rg#2,\end,}}
\def\first@rg#1,{\callr@nge{#1}\apply@rg}
\def\apply@rg#1,{\ifx\end#1\let\next=\relax%
\else,\callr@nge{#1}\let\next=\apply@rg\fi\next}

\def\callr@nge#1{\calldor@nge#1-\end-}
\def\callr@ngeat#1\end-{#1}
\def\calldor@nge#1-#2-{\ifx\end#2\thecsname#1 ,%
  \else\thecsname#1 ,--\thecsname#2 ,\callr@ngeat\fi}

\def\writenew#1#2{\immediate\write\xrffile
     {\string\def\csname @@#1num#2 \endcsname{\t@ghead\number\t@g}}}

\def\ifunc@lled#1#2{\expandafter\ifx\csname #1num#2\endcsname\relax}

\def\t@ghead{}
\newcount\t@ghd\t@ghd=0
\def\taghead#1{\gdef\t@ghead{#1}\global\advance\t@ghd by 1}

\ifx\eqn\undefined
  \order{eqn}\let\@qn=\eqn
\else
  \let\eqn@=\eqn\order{eqn}\let\@qn=\eqn\let\eqn=\eqn@
\fi\let\ref@qn=\refeqn

\let\eqno@=\eqno
\def\eqno(#1){\eqno@({\rm\@qn{#1}})}

\ifx\eq\undefined
  \let\eq@=\relax
\else
  \let\eq@=\eq
\fi
\def\eq(#1){\eq@({\rm\@qn{#1}})}

\def\refeq#1{{\rm(\ref@qn{#1})}}

\newcount\r@fcount \r@fcount=0
\def\refcite#1{\each@rg\r@featspace{#1}}
\def\r@featspace#1#2 #3{\r@fcite#1#2,}
\def\r@fcite#1,%
  {\ifunc@lled{r@f}{#1}\global\advance\r@fcount by 1%
    \expandafter\xdef\csname r@fnum#1\endcsname{\number\r@fcount}%
    \expandafter\gdef\csname r@ftext\number\r@fcount\endcsname%
     {\message{ Reference #1 to be supplied }
      \if@iopp\refjl{}{}{}{Reference #1 to be supplied} \par
      \else Reference #1 to be supplied\par\fi}%
   \fi\csname r@fnum#1\endcsname}

\def\refis#1 #2\par
  {\ifunc@lled{r@f}{#1}
  \else
    \global\let\refjl=\numrefjl\global\let\refbk=\numrefbk
    \expandafter\gdef\csname r@ftext\csname r@fnum#1\endcsname\endcsname%
    {\writenewr@f#1>>#2\par}
  \fi}

\def\writenewr@f#1>>{}
\gdef\referencefile{\expandafter\immediate\csname newwrite\endcsname\reffile
                    \immediate\openout\reffile=\jobname.ref
   \def\writenewr@f##1>>%
  {\immediate\write\reffile{\noexpand\refis{##1}
     \expandafter\expandafter\expandafter\strip@t\expandafter
     \meaning\csname r@ftext\csname r@fnum##1\endcsname\endcsname}}
  \def\strip@t##1>>{}}

\newcount\r@fcurr
\def\listreferences{\global\r@fcurr=0
  {\loop\ifnum\r@fcurr<\r@fcount\global\advance\r@fcurr by 1
   \if@iopp\let\refjl=\numr@@fjl\let\refbk=\numr@@fbk
     \csname r@ftext\number\r@fcurr\endcsname
   \else
     \numr@f{\number\r@fcurr}{\csname r@ftext\number\r@fcurr\endcsname}
   \fi\repeat}}

\newif\if@iopp
\ifx\numrefjl\undefined
  \@ioppfalse
\else
  \ifx\numrefbk\undefined
    \@ioppfalse
  \else
    \@iopptrue
  \fi
\fi

\if@iopp
  \message{ IOPP reference macros in use }
  \let\r@fjl=\refjl
  \let\numr@fjl=\numrefjl
  \let\r@fbk=\refbk
  \let\numr@fbk=\numrefbk
  \def\numr@@fjl#1#2#3#4{\numr@fjl{\number\r@fcurr}{#1}{#2}{#3}{#4}
     \let\refjl=\r@fjl\let\refbk=\r@fbk}
  \def\numr@@fbk#1#2#3{\numr@fbk{\number\r@fcurr}{#1}{#2}{#3}
     \let\refjl=\r@fjl\let\refbk=\r@fbk}
\else
  \message{ User defined reference macros }
  \def\numr@f#1#2 {\parindent=30pt
  \hang\noindent\rm {\hbox to 30truept{[#1]\hss\quad}}#2
  \par\parindent=16pt}
\fi

\catcode`@=12

\order{thm}\order{lem}

\def\be{$$}
\def\ee#1{\eqno(#1)$$}                            

\def\summe#1#2{\sum_{#1=1}^{#2}}
\def\produkt#1#2{\prod_{#1=1}^{#2}}
\def\sumexc#1#2#3{\sum_{#1=1\atop #1\not= #2}^{#3}}
\def\cre#1#2{c^+_{#1,#2}}
\def\ann#1#2{c_{#1,#2}}
\def\num#1#2{n_{#1,#2}}
\def\Msum#1#2{\sum_{#1=1}^{M_{#2}}}
\def\Mprod#1#2{\prod_{#1=1}^{M_{#2}}}
\def\nusum#1#2{\sum_{#1=1}^{\nu_{#2}}}
\def\Lam#1#2{\Lambda_{#1}^{(#2)}}
\def\th#1#2{\vartheta_{#1}^{(#2)}}
\def\N{{\cal N}}
\def\K{K}
\def\kint{\int_{-\K}^\K}
\def\j{j}
\def\jr{j}
\def\e{{\rm e}}
\def\O{O}
\def\qLam#1#2{q^{#1}\left({i\over 2}#2 \eta\right)}
\def\qLam2#1#2{q^{#1}\left({i\over 2}#2 {\eta\over 2}\right)}
\def\etah{{\eta\over 2}}
\def\ih{{i\over 2}}
\def\T{T}
\def\t{t}

\title {Excitation spectrum and critical exponents of a one-dimensional
integrable model of fermions with correlated hopping\footnote*{Work
performed within the research program of the Sonderforschungsbereich\ 341,
K\"oln-Aachen-J\"ulich}}

\author{R.\ Z.\ Bariev\dag , A.\ Kl\"umper ,
A.\ Schadschneider , and J.\ Zittartz}

\address{Institut f\"ur Theoretische Physik, Universit\"at zu K\"oln,
Z\"ulpicher Stra{\ss}e 77, D-5000 K\"oln, Federal
Republic of Germany}

\address{\dag\ Permanent address: The Kazan Physico-Technical Institute
of the Russian Academy of Sciences, Kazan 420029}

\vfill

\shorttitle{Spectrum of fermions with correlated hopping}


\jnl{J. Phys. A}

\date

\beginabstract
We investigate the excitation spectrum of a model of $N$ colour
fermions with correlated hopping which can be solved by a nested
Bethe ansatz. The gapless excitations
of particle-hole type are calculated as well as the spin-wave like
excitations which have a gap.
Using general predictions of conformal field theory
the long distance behaviour of some groundstate correlation functions are
derived from a finite-size analysis of the gapless excitations.
{}From the algebraic decay we show that for increasing particle density
the correlation of so-called $N$-multiplets of particles dominates over the
density-density correlation. This indicates the presence of bound complexes
of these $N$-multiplets. This picture is also supported by the calculation
of the effective mass of charge carriers.
\endabstract

\section{Introduction}
The study of low-dimensional electronic systems with strong correlations
has gained considerable importance due to the discovery of high-temperature
superconductivity. The models which have been mostly studied
are the one-dimensional Hubbard model [1]
and the $t-J$ model [2,3] at its supersymmetric
point. These models are special because of their
integrability [4-7].
A great deal of current research is directed to more extensive
rigorous and exact results on correlated systems in one and two dimensions.
However, in the latter case exact results are sparse. In contrast to this
we have a different situation in one dimension where many models
are integrable and show different physical behaviour.

In this paper we continue the study of a fermionic model with correlated
hopping [8]. In addition to on-site Coulomb interactions
as in the above mentioned cases we are interested in
modifications of the hopping terms of fermions in the vicinity of other
particles.
Unfortunately, such complex models are not integrable in general. However,
for purely correlated hopping the situation looks much better. In [9] an
integrable model of electrons was found which was generalized to fermions
with arbitrary number
$N\ge 2$ of colours [10]. The Hamiltonian of this model is
given by
$$
H=-\summe{j}{L}
\summe{\tau}{N}
\bigl(\cre{j}{\tau}\ann{j+1}{\tau}+\cre{j+1}{\tau}\ann{j}{\tau}\bigr)
\cdot\exp\Biggl(-\eta\sumexc{\tau'}{\tau}{N}
\num{j+\theta(\tau-\tau')}{\tau'}
\Biggr),
\eqno(hamil)
$$
where $L$ is the length of the chain,
$\cre{j}{\tau}$ and $\ann{j}{\tau}$ are the creation and annihilation
operators of fermions of colour $\tau$ at site $j$ ($\tau=1,...,N$ and
$j=1,...,L$), $\num{j}{\tau}$ is the number operator and
$\theta$ is the step function
\be
\theta(\tau-\tau')=\cases{1,&$\tau>\tau'$,\cr
                          0,&$\tau<\tau'$.}
\ee{defth}
We employ the usual periodic boundary conditions
$\ann{L+1}{\tau}=\ann{1}{\tau}$.
Repulsion of particles corresponds to an interaction parameter $\eta>0$.
In [8] the case $N=2$ of model \refeq{hamil} was investigated, however for
negative values of $\eta$. Note that $\eta$ and $-\eta$ are related by a
particle-hole transformation.
Using a Jordan-Wigner transformation the model can be formulated
equivalently as a system of $N$ interacting $XY$ chains [10].

In Sect. 2 we give a summary of the Bethe Ansatz equations and some groundstate
properties. In Sect. 3 we study the excitations of the model.
In Sect. 4 we derive
the long-distance behaviour of correlation functions and compare the
results with previous ones  [8] for the special case of $N=2$.
In the Appendix a short-cut to the analysis of Bethe Ansatz equations is
presented on the basis of inversion identities.

\section{The Bethe Ansatz}
The Bethe Ansatz for this model has been derived in [10] from which we quote
the
relevant equations. The eigenstates of the Hamiltonian are characterized by
sets of wavenumbers $k_j$, $j=1,...,\N$, for $\N$ particles.
There are additional Bethe Ansatz parameters $\Lam{\alpha}{r}$
($r=1,...,N-1$ and $\alpha=1,...,M_{N-r}$) the number of which
is given by
\be
M_{N-r}=\summe{j}{N-r}\N_j,
\ee{defM}
where $\N_j$ denotes the number of particles of colour $j$.
The parameters $k_j$ and $\Lam{\alpha}{r}$ satisfy the set of nested Bethe
Ansatz equations derived in [10]
$$
\eqalign{
&k_j L+\Msum{\alpha}{N-1}\Theta\Bigl(k_j-\Lam{\alpha}{1};{\eta\over 2}\Bigr)
=2\pi I_j,\cr
&\sum_{\sigma=\pm 1}\Msum{\beta}{N-r-\sigma}
\Theta\Bigl(\Lam{\alpha}{r}-\Lam{\beta}{r+\sigma};{\eta\over 2}\Bigr)
-\Msum{\beta}{N-r}
\Theta\Bigl(\Lam{\alpha}{r}-\Lam{\beta}{r};{\eta}\Bigr)
=2\pi J_\alpha^{(N-r)},}
\eqno(BAGl)
$$
with $j=1,...,\N$; $r=1,...,N-1$; $\alpha=1,...,M_{N-r}$, and we have set
$\Lam{j}{0}=k_j$. The phase shift function $\Theta$ is given by
\be
\Theta(k;\eta)=2\arctan\left(\coth\eta\tan{1\over 2}k\right),
\quad -\pi\le\Theta<\pi,
\ee{defTh}
and $I_j$ and $J_\alpha^{(r)}$ are integer (half-integer) numbers for
odd (even) $M_{N-1}+1$ and $M_{r-1}+\N_{r+1}$, respectively.
Energy and momentum of the corresponding state are given by
$$
\eqalign{
E&=-2\summe{j}{\N}\cos k_j,\cr
P&=\summe{j}{\N}k_j={2\pi\over L}\Biggl(\summe{j}{\N}I_j
+\summe{r}{N-1}\Msum{\alpha}{N-r}J_\alpha^{(N-r)}\Biggr).\cr
}\eqno(Energie)
$$

It follows by symmetry that the ground state
corresponds to symmetric configurations with the same number of particles
$\N_i$ for all colours $i$
\be
\N_i={\N\over N}, \qquad 1\le i\le N.
\ee{GZ}
The ground state energy is calculated from \refeq{BAGl}--\refeq{Energie}
in the thermodynamic limit [10]
\be
{E_0\over L}=-2\kint \cos k \,\rho(k)dk,
\ee{GZenergie}
where the density function $\rho(k)$ is determined as the solution of the
integral equation
$$
\eqalign{
&2\pi\rho(k)-\kint\varphi(k-k')\rho(k')dk'=1,\cr
&\varphi(k)=1-{1\over N}+2\summe{n}{\infty}\exp(-n\eta)
{\sinh\bigl[n\eta(N-1)\bigr]\over\sinh(n\eta N)}\cos(nk),
}\eqno(GZgl)
$$
and $K$ is determined by the subsidiary condition
\be
\kint\rho(k)dk={\N\over L}=\rho.
\ee{subsi}
Equations \refeq{GZenergie}--\refeq{subsi} determine the ground state
energy of the model as a function of the particle density $\rho$.
The Appendix contains a derivation of eqs. \refeq{GZenergie}--\refeq{subsi}
from (4)-(6) which is slightly different from the original treatment in [10].

In \refeq{GZ} we have assumed $\N$ a multiple of $N$. This in
fact is the condition for a proper accomodation of the `antiferromagnetic'
ground state. For particle numbers $\N$ not multiples of $N$ we
always have a misfit resulting in `spin-type' excitations which are
treated below. This kind of excitations has a non-zero gap by which the
bulk energies of the ground states of systems with $\N$ not multiples of
$N$ are increased.

\section{Excited states}
The system of Bethe Ansatz equations \refeq{BAGl} admits many solutions
depending on the choice of the parameters $I_j$ and $J_\alpha^{(r)}$.
In particular the solution for the ground state \refeq{GZenergie} is
characterized by a parameter set in which the $I_j$ and $J_\alpha^{(r)}$
are consecutive integers (or half-integers) centered around the origin.
It is expected that the low-lying states are obtained by small modifications
to these configurations.

There are two types of elementary excitations leaving the particle
number unchanged. The first one is of particle-hole type corresponding
to raising one pseudo-particle in the ground state $|k_h|<\K$ to a higher
level $|k_p|>\K$. This excitation is obtained by a straight forward
manipulation of the set of numbers $I_j$ without changing the parameters
$J_\alpha^{(r)}$. This produces gapless excitations analogous to those in
the antiferromagnetic spin-1/2 Heisenberg chain [11,12] and in the repulsive
Hubbard model [13-15]. In the thermodynamic limit energy and momentum of these
excitations are given by
$$
\eqalign{
E-E_0&=\summe{j}{\nu}\epsilon\bigl(k_j^p)
-\summe{j}{\nu}\epsilon\bigl(k_j^h\bigr),\cr
P-P_0&=\summe{j}{\nu}p\bigl(k_j^p)-\summe{j}{\nu}p\bigl(k_j^h\bigr),
}\eqno(TeilLoch)
$$
with energy-momentum dispersion of particle-hole excitations
$$
\eqalign{
\epsilon\bigl(\vartheta\bigr)&=-2\Biggl[\cos \vartheta
+\kint\sin k\, \j(k) dk\Biggr],\cr
p\bigl(\vartheta\bigr)&=\vartheta - \kint\j(k)dk.\cr
}\eqno(DispTeilLoch)
$$
The function $\j(k)$ is the solution of the integral equation
\be
2\pi\j(k)-\kint\varphi(k-k')\j(k')dk'=\Phi(k-\vartheta),
\ee{TeilLochGl}
where $\Phi'(k)=\varphi(k)$. Alternatively, the dressed energy function
$\epsilon$ can be obtained directly from an integral equation
identical to the last one upon the replacement of $\j(k)$ and
$\Phi(k-\vartheta)$ by $\epsilon(k)$ and $\epsilon_0(k)=-2\cos k$,
respectively. In general \refeq{TeilLochGl} can be solved
only numerically. In certain
limiting cases also analytic results can be obtained. For instance
in the strong-coupling limit ($\eta\to\infty$) perturbation theory
yields
$$
\eqalign{
\epsilon\bigl(\vartheta\bigr)=&
-2\Biggl[1+{\e^{-2\eta}\over\pi}\Bigl(\K-{1\over 2}\sin 2\K\Bigr)\Biggr]
\Bigl(\cos
\vartheta+{t\over\pi}(\sin\K-\K\cos\K)\Bigr)+O\bigl(\e^{-2\eta}\bigr),\cr
p\bigl(\vartheta\bigr)=&{\pi\over\pi-\K t}\vartheta
+{2\e^{-2\eta}\sin\K\over\pi-\K t}
\Biggl[\sin \vartheta+{t\sin\K\over\pi-\K t}\vartheta\Biggr]
+O\bigl(\e^{-4\eta}\bigr),\qquad t={N-1\over N}.\cr
}\eqno(grenzfall)
$$

The excitations of the second type involve ``$r$-holes" and ``strings"
in the sets $\Lam{\alpha}{r}$. A ``$r$-hole" occurs when
there is a jump in the sequence of $J_\alpha^{(r)}$. Physically it describes
states where the colour of one bare particle has been changed.
In a sense these excitations are of spin-wave type, but have a
non-zero gap. The general energy-momentum excitation
is given by $N-1$ different dispersion functions
$$
\eqalign{
E-E_0=\summe{r}{N-1}\nusum{j}{N-r}\epsilon^{(N-r)}\Bigl(\th{j}{r}\Bigr),\cr
P-P_0=\summe{r}{N-1}\nusum{j}{N-r}p^{(N-r)}\Bigl(\th{j}{r}\Bigr).
}\eqno(Spinanreg)
$$
Each elementary excitation parametrized with a rapidity variable $\vartheta$
is given by
$$
\eqalign{
\epsilon^{(N-r)}(\vartheta)&=2\kint\sin k \,\jr(k)dk,\cr
p^{(N-r)}(\vartheta)&=\kint\jr(k)dk,\cr
}\eqno(SpinanregDisp)
$$
where $\jr(k)$ is the solution of an integral equation of the type
\refeq{TeilLochGl}, however with different right hand side
\be
2\pi\jr(k)-\kint\varphi(k-k')\jr(k')dk'=\varphi^{(N-r)}(k-\vartheta),
\ee{SpinanregGl}
where
\be
\varphi^{(N-r)}(k)={N-r\over N}k+2\summe{n}{\infty}
{\sinh\bigl[n\eta(N-r)\bigr]\over\sinh(n\eta N)}{\sin(nk)\over n}.
\ee{inh}
The lowest energies $\Delta^{(r)}$ of \refeq{SpinanregDisp} are
taken at $\vartheta=\pi$. In Fig. 1 the gaps are shown for different
values of $N$ and $\eta$ in dependence of the particle density $\rho$.
Not all combinations of the elementary excitations \refeq{SpinanregDisp}
are allowed, the numbers $\nu_r$
in \refeq{Spinanreg} have to satisfy a selection rule
\be
\summe{r}{N-1}r\nu_{N-r}\equiv 0 {\pmod N}.
\ee{Auswahl}

For a derivation of \refeq{SpinanregGl} and \refeq{inh} utilizing
inversion identities see the Appendix.
In general \refeq{SpinanregDisp}, \refeq{SpinanregGl} have to be
solved numerically.
In the strong-coupling limit
($\eta\to\infty$) the analytic solution of
\refeq{SpinanregDisp}--\refeq{inh} is
$$
\eqalign{
\epsilon^{(N-r)}(\vartheta)&=\Delta^{(N-r)}+{2\e^{-r\eta}\over\pi}
\Bigl(\K-{1\over 2}\sin 2\K\Bigr)(1+\cos\vartheta)
+O\Bigl(\e^{-(r+2)\eta}\Bigr),\cr
p^{(N-r)}(\vartheta)&={\K\over\pi-\K t}{N-r\over N}(\pi-\vartheta)
-{2T_1\sin\K\over\pi-\K t}+O\Bigl(\e^{-4\eta}\Bigr),\cr
T_1&=\e^{-r\eta}\sin\vartheta
-\e^{-2\eta}{N-r\over N}{\pi-\vartheta\over\pi-\K t}\sin\K,\cr
\Delta^{(N-r)}&=2T_2\Biggl[1+{\e^{-2\eta}\over\pi}
\Bigl(\K-{1\over 2}\sin
2\K\Bigr)\Bigl(1-T_2^{-1}\e^{(2-r)\eta}\Bigr)\Biggr],\cr
T_2&={N-r\over\pi N}\bigl(\sin\K-\K\cos\K\bigr).\cr
}\eqno(Limes)
$$

In addition to simple holes excited states may also contain
conjugate pairs of complex $\Lam{\alpha}{r}$. The additional ``strings"
do not contribute to energy and momentum which effect
is very transparent in the
inversion identity approach, see the Appendix.
Such a cancellation phenomenon is well-known for the Heisenberg model,
see for instance [12].
Concerning the spectrum of the model these states only lead to a
degeneracy of the energy levels. With respect to the classification
of all eigenstates, however, they do play an important role.

Thus in the considered model there are no gapless excitations of spin
wave type in contrast to the antiferromagnetic Heisenberg chain [11-12] and
the repulsive Hubbard model [13-15]. Instead the analogous excitations
correspond to collective motions of pseudo-particles which can be
regarded as new particles with non-zero masses.

The gap implies the binding of particles in multiplets of $N$ particles
with different colours. Any spatial separation of the constituents of
these complexes would lead to two subsystems of the chain with particle
numbers not multiples of $N$. As explained in Sect. 2 this results in
an increase of the ground state energies by the order of the ``spin-gap",
and eventually leads to the exponential decay of the corresponding
correlation function.

\section{Critical exponents of the correlation functions}
In the Bethe Ansatz approach it is a formidable task to deal with
correlation functions. However, due to developments in two-dimensional
conformal field theory the scaling dimensions describing the algebraic
decay of correlation functions became accessible [16,17]. According
to this theory there is a one-to-one correspondence between the scaling
dimensions and the spectrum of the quantum system on a (periodically
closed) finite chain [18,19]
$$
\eqalign{
E_j-E_0&={2\pi v_F\over L}\Bigl(x_j+N^+ + N^-\Bigr),\cr
P_j-P_0&={2\pi \over L}\Bigl(s_j+N^+ - N^-\Bigr)+2dk_F.\cr
}\eqno(Skalen)
$$
$E_j$ and $P_j$ are the energy and momentum of the $j$th
excited state, $v_F$ is the Fermi velocity. The $x_j$ and $s_j$ are
the scaling dimensions and ``spins" of the corresponding operators,
$N^+$, $N^-$ are non-negative integers and $d$ is the number of
particles excited from the left Fermi point to the right one.
The ground state energy is expected to scale like [20,21]
\be
E_0=L\epsilon_\infty-{\pi v_F \over 6L}c,
\ee{Zentral}
where $\epsilon_\infty$ is the ground state energy per site of the infinite
system and $c$ is the central charge characterizing the underlying
conformal field theory.

Consequently the scaling dimensions are obtained from the gaps due to
finite-size effects in the spectrum of the Hamiltonian at criticality.
In order to compute the finite-size corrections we use a method
which has been developed by previous authors for the Heisenberg and
Hubbard chains [22,23]. As a result we have obtained the ground state energy
\refeq{Zentral} with central charge $c=1$ and the finite-size
spectrum of the low-lying excitations
$$
\eqalign{
E-E_0&={2\pi v_F\over L}\left( \Biggl[{\Delta \N\over 2\xi(\K)}\Biggr]^2
+[\xi(\K)]^2d^2+N^++N^-\right),\cr
P&=2 k_Fd+{2\pi\over L}\left(d\Delta \N +N^+-N^-\right), \qquad
k_F=\pi\rho.\cr}
\eqno(spektrum)
$$
Here the dressed charge at the Fermi surface $\xi(\K)$ is related in a
simple way to the function $\rho(k)$ in \refeq{GZgl}
\be
\xi(\K)=2\pi\rho(\K),
\ee{dressCharge}
and $\Delta\N$ is the change of the particle number compared to the ground
state. The non-negative integers $N^\pm$ describe the excitations of
particle-hole type in the vicinity of the Fermi points $\pm k_F$. Now we
can read off all critical exponents from \refeq{spektrum} and
\refeq{Skalen}
$$
\eqalign{
x_{\Delta\N,d}&={(\Delta \N)^2\over [2\xi(\K)]^2} + [\xi(\K)]^2 d^2
+N^++N^-,\cr
s_{\Delta\N,d}&=d\Delta\N+N^+-N^-,\cr
}\eqno(xunds)
$$
where $\Delta\N$ and $d$ can take the values
$$
\eqalign{
\Delta\N&=0,\, N,\, 2N,\, ...,\cr
d&\equiv{\Delta\N\over 2N} {\pmod {{1\over N}}},\cr
}\eqno(nebenbed)
$$
generalizing the result in [24] for $N=2$.
The omitted integer values of $\Delta\N$ correspond to
excitations with gap which are not related to correlations with
algebraic decay. The two-point correlation functions of the scaling
fields $\Phi_{\Delta^\pm}(x,t)$ with conformal weights $\Delta^\pm
=(x\pm s)/2$ are known to be [16,19]
\be
\langle\Phi_{\Delta^\pm}(x,t)\Phi_{\Delta^\pm}(0,0)\rangle=
{\exp(-2idk_Fx)\over(x-v_Ft)^{2\Delta^+}(x+v_Ft)^{2\Delta^-}}.
\ee{allgKorr}
The dimensions of the descendant fields differ from $\Delta^\pm$ by the
integers $N^\pm$.

Let us first consider the density-density correlation function. In this case
the correlation function is determined by excitations with unchanged
number of particles $\Delta\N=0$. The leading contributions to the algebraic
decay of the correlation are then given by $d=0$,
$(N^+,N^-)=(1,0)$ or $(0,1)$, and by $d=1/N$, $N^\pm=0$, respectively.
This leads to the asymptotic form
\be
\langle \rho(r)\rho(0)\rangle \simeq \rho_0^2+A_1r^{-2}+A_2r^{-\alpha}
\cos\left({2\over N}k_Fr\right),
\ee{DichteKorr}
where
$$
\rho(r)=\summe{\tau}{N}\num{r}{\tau},\qquad
\alpha={2\xi(\K)^2\over N^2}.\eqno(defDichte)
$$

Field-field correlators decay exponentially as the corresponding
excitations change the particle number $\N$ by one unit, as remarked before.
However, we may consider the correlation function of a $N$-multiplet of fields,
\be
\O(r)=\produkt{\tau}{N}\ann{r}{\tau}\,.
\ee{defO}
In this case the particle number changes by
$\Delta\N=N$ which according to (26) is the smallest non-zero
value for which we have algebraic decay. We have $d=0$ or $1/2N$
for even or odd $N$, respectively.
We thus obtain
$$
\langle \O^+(r)\O(0)\rangle \simeq r^{-\beta}\cdot\cases
{1,&\cr
\cos\left({k_F\over N} r\right),&}
\qquad\beta=\cases
{1/\alpha,&$N$ even,\cr
1/\alpha+\alpha/4,&$N$ odd.}
\eqno(KorrO)
$$
The characteristic exponent (29) is determined numerically and plotted
in Fig. 2 for the cases $N=3$, 4. For
$N=2$ we refer to [8].
Analytically we find $\alpha(\rho=0)=2/N^2$ and $\alpha(\rho=N)=2$.
This shows that for all $N$ there exists an interval $[\rho_c,N]$ for
which $\beta < \alpha$. In this case the $N$-multiplet correlation (31) has a
slower decay than the density correlation (28) and thus dominates. This might
indicate the presence of bound complexes of $N$-multiplets in generalization
of bound pairs for $N=2$ which has been discussed in [8].

In order to get a more complete picture of the physical properties of the model
we also investigate the conductivity and the effective transport mass.
Following
the ideas of [25,26] we study the Hamiltonian \refeq{hamil} with twisted
boundary conditions with twisting angle $\varphi$. In this case the Bethe
Ansatz equations
read
$$
\eqalign{
&k_j L+\Msum{\alpha}{N-1}\Theta\Bigl(k_j-\Lam{\alpha}{1};{\eta\over 2}
\Bigr)=2\pi I_j+\varphi,\cr
&\sum_{\sigma=\pm 1}\Msum{\beta}{N-r-\sigma}
\Theta\Bigl(\Lam{\alpha}{r}-\Lam{\beta}{r+\sigma};{\eta\over 2}\Bigr)
-\Msum{\beta}{N-r}
\Theta\Bigl(\Lam{\alpha}{r}-\Lam{\beta}{r};{\eta}\Bigr)
=2\pi J_\alpha^{(N-r)}.}
\eqno(TBAG)
$$
The additional phase $\varphi$ corresponds to an enclosed magnetic flux in the
ring.

The conductivity is directly proportional to the charge stiffness (Drude
weight)
$D_c$ which can be obtained from the change
$$
\Delta E_0=D_c\varphi^2/L
\eqno(ETBC)
$$
in the ground state energy for small $\varphi$ [25]. Using the charge stiffness
we can define an effective transport mass $m$ by the relation
$$
{m\over m_e}={D_c^0\over D_c}
\eqno(MASSDEF)
$$
where $D_c^0={N\over \pi}\sin\left({\pi \rho\over N}\right)$ is the charge
stiffness
of the non-interacting ($\eta=0$) system and $m_e$ is the (bare) electron mass.

Using the result \refeq{spektrum} for the finite-size corrections with
$\Delta {\cal N} = 0$ and $d=\varphi/2\pi$ we find
$$
D_c={1\over 2\pi} v_F\xi^2(K)={1 \over 2\pi}\epsilon'(K)\xi(K).
\eqno(DC)
$$
Using the integral equation \refeq{GZgl} and \refeq{dressCharge}
we can calculate $D_c$ and $m$ numerically for all densities $\rho$, see
Fig.~3.
For large particle densities $\rho$ and strong interaction $\eta$ we
observe an increase of the mass $m$ by a factor $N$ as compared to the
non-interacting case. This is consistent with and supports our above
picture that bound complexes of $N$-multiplets are present in the system.

\appendix{}

Here we describe the reduction of the Bethe Ansatz problem \refeq{BAGl}
involving $k_j$ and $\Lam{\alpha}{r}$ parameters to equations involving
only the $k_j$ variables. To this end we cite a set of equations which
are equivalent to \refeq{BAGl} and which appeared at an intermediate
stage of the analysis in [10]
\be
\e^{ik_j L}=\T^1\left({i\over 2}k_j+\etah\right).
\ee{Randbdg}
$\T^1(u)$ is the first of a set of functions $\T^1(u)$, $\T^2(u)$,...,
$\T^{N-1}(u)$,
satisfying the recursion
\be
\T^r\left(u\right)={1\over q^r(u)}\left[q^r(u-\eta)+q^r(u+\eta)
{q^{r-1}\left(u-\etah\right)\over q^{r-1}\left(u+\etah\right)}
\T^{r+1}\left(u+\etah\right)\right],
\ee{Rekursion}
with $\T^N(u)\equiv 1$. The $q$-functions are given by
$$
q^r(u)=\Mprod{\alpha}{N-r}\sinh\left(u-\ih \Lam{\alpha}{r}\right),
\eqno(Defq)
$$
in terms of the parameters $\Lam{\alpha}{r}$ which are subject to
the condition that all $\T^r(u)$ be analytic functions with simple poles at
$\ih\Lam{\alpha}{r-1}-\etah$. This condition is equivalent to the set
of Bethe ansatz equations \refeq{BAGl}. Note that $\Lam{0}{r}=k_j$ and
$M_N=\N$.

On the right hand side of \refeq{Rekursion} one of the
two summands usually dominates over the other. This provides the possibility
to derive a simple functional equation for $\T^1(u)$ in product form
\be
\T^1(u)\T^1(u+\eta)...\T^1(u+(N-1)\eta)\simeq{q^0\left(u-\etah\right)\over
q^0\left(u+\left(N-{3\over 2}\right)\etah\right)},
\ee{InvRel}
where the right hand side depends only on the $k_j$ parameters
and the corrections to this relation are exponentially small in the
thermodynamic limit. For the
ground state \refeq{InvRel} can be solved uniquely (apart from $N$th roots
of unity) under the condition
of analyticity, absence of zeros in the physical regime
$-(N-1)\eta<$Re$(u)<\eta$, and the pole structure at $\ih k_j-\etah$.
The result can be written in the form
\be
\ln\T^1\left(\ih k+\etah\right)=i{2\pi\over N}j+
{1\over i}\summe{l}{\N}\Phi(k-k_l),\qquad j=0,1,...,N,
\ee{Loesung}
where
\be
\Phi(k)=\left(1-{1\over N}\right)(k-\pi)+2\summe{n}{\infty}
{\exp(-n\eta)\over n}{\sinh\bigl[n\eta(N-1)\bigr]\over\sinh(n\eta N)}\sin(nk).
\ee{psi}
Using standard reasoning from this and \refeq{Randbdg} one derives
the integral equation \refeq{GZgl}. Note that $\varphi(k)=\Phi'(k)$. The
absolute ground state of the system corresponds to the choice of $j=0$ in
\refeq{Loesung}. Other choices of $j$ lead to excitations with energies of
order $1/L$ and give rise to a quantization of the $d$ parameter in
\refeq{nebenbed} in units of $1/N$.

For excitations of spin-wave type it is useful to define the excitation
function
\be
\t(u)=\lim_{L\to\infty}{\T^1(u)\over\T^1_{\rm max}(u)},
\ee{Anregfkt}
for which \refeq{InvRel} directly implies
\be
\t(u)\t(u+\eta)...\t(u+(N-1)\eta)=1.
\ee{invrel}
A first consequence of this functional equation is the double periodicity
of the analytic continuation of $\t(u)$
$$
\eqalign{
&\t(u)=\t(u+i\pi),\cr
&\t(u)=\left[\t(u+\eta)\t(u+2\eta)...\t(u+(N-1)\eta)\right]^{-1}
=\t(u+N\eta).\cr
}\eqno(DoppelPer)
$$
The elementary spin-wave excitations are characterized by pairs of
zeros and poles which have to appear in pairs $u_z$ and $u_p$. The distance
of $u_z$ and $u_p$ has to be a multiple of $\eta$ in order to lead to the
cancellation \refeq{invrel}. We write
\be
u_{z,p}={1\over 2}\left(\eta+i\vartheta\mp r\etah\right),\qquad r=1,...,N-1,
\ee{Sing}
where $\vartheta$ is a rapidity variable and takes real numbers.
The solution of \refeq{invrel} under the condition of analyticity and
the zero/pole structure of \refeq{Sing} is
\be
\ln\t\left(\ih k+\etah\right)=i\varphi^{(N-r)}(k-\vartheta),
\ee{loesung}
where the function on the right hand side was defined already in \refeq{inh}.
Again by standard reasoning and \refeq{Randbdg} one finds the integral
equation \refeq{SpinanregGl}. It is noteworthy that in this approach the
distribution of $\Lambda$-parameters is quite irrelevant. Different patterns,
in particular simple holes with or without strings, lead to the same function
\refeq{loesung}.

\references

\numrefjl{[1]}{J.\ Hubbard 1963}{\PRS London A}{276}{238}
\numrefjl{[2]}{P.\ W.\ Anderson 1987}{Science}{235}{1196 }
\numrefjl{[3]}{F.\ C.\ Zhang and T.\ M.\ Rice 1988}{\PRB}{37}{3759 }
\numrefjl{[4]}{E.\ H.\ Lieb and F.\ Y.\ Wu 1968}{\PRL}{20}{1445 }
\numrefjl{[5]}{B.\ Sutherland 1975}{\PRB}{12}{3795 }
\numrefjl{[6]}{P.\ Schlottmann 1987}{\PRB}{36}{5177 }
\numrefjl{[7]}{P.\ A.\ Bares and G.\ Blatter 1990}{\PRL}{64}{2567 }
\numrefjl{[8]}{R.\ Z.\ Bariev, A. Kl\"umper, A. Schadschneider and
J. Zittartz 1993}{\JPA}{26}{1249}
\numrefjl{[9]}{R.\ Z.\ Bariev 1991}{\JPA}{24}{L549}
\numrefjl{[10]}{R.\ Z.\ Bariev 1991}{\JPA}{24}{L919}
\numrefbk{[11]}{M.\ Gaudin}{La Fonction d'Onde Bethe}{(Paris, Masson) 1983}
\numrefjl{[12]}{L.\ D.\ Faddeev and L.\ A.\ Takhtajan 1981}{\PLA}{85}{375}
\numrefjl{[13]}{A.\ A.\ Ovchinnikov 1970}{Sov. Phys. \JETP}{30}{1160}
\numrefjl{[14]}{C.\ F.\ Coll 1974}{\PRB}{9}{2150}
\numrefjl{[15]}{A. Kl\"umper, A. Schadschneider and J. Zittartz 1990}
{\ZPB}{78}{99}
\numrefjl{[16]}{A.\ A.\ Belavin, A.\ M.\ Polyakov and A.\ B.\ Zamolodchikov
1984}{Nucl.\ Phys.\ B}{241}{333 }
\numrefjl{[17]}{J.\ L.\ Cardy 1986}{Nucl.\ Phys.\ B}{270 [FS16]}{186 }
\numrefjl{[18]}{J.\ L.\ Cardy 1984}{\JPA}{17}{L385}
\numrefjl{[19]}{N.\ M.\ Bogoliubov, A.\ G.\ Izergin and N.\ Yu.\ Reshetikhin
1987}{\JPA}{20}{5631}
\numrefjl{[20]}{H.\ W.\ J.\ Bl\"ote, J.\ L.\ Cardy and M.\ P.\ Nightingale
1986}{\PRL}{56}{742}
\numrefjl{[21]}{I.\ Affleck 1986}{\PRL}{56}{746}
\numrefjl{[22]}{F.\ Woynarovich, H.\ P.\ Eckle and T.\ T.\ Truong 1989}
{\JPA}{22}{4027}
\numrefjl{[23]}{F.\ Woynarovich 1989}{\JPA}{22}{4243 }
\numrefjl{[24]}{R.\ Z.\ Bariev 1991}{\JPA}{25}{L 261}
\numrefjl{[25]}{B.\ S.\ Shastry and B.\ Sutherland 1990}{\PRL}{66}{243 }
\numrefjl{[26]}{N.\ Kawakami and S.-K.\ Yang 1991}{\PRB}{44}{7844 }

\figures
\figcaption{(a) Depiction of the gap of the massive excitations
in dependence of the particle density $\rho$ for $N=2$ and
different interaction strengths $\eta=$ 0.5, 1, 2, $\infty$.
(b) The excitation gap for the four massive dispersions in the case
$N=5$ and $\eta=1$.}

\figcaption{(a) Depiction of the critical exponent $\alpha$ for $N=3$ and
different values of $\eta=$ 0, 0.1, 0.5 $\infty$. For values of $\alpha$ above
the broken line we have dominating $N$-multiplet correlations.
(b) The same for $N=4$.}

\figcaption{The effective mass of the charge carriers for $N=3$ and
different interaction strengths $\eta=$ 0, 0.1, 0.5, 1, 10. For
particle densities close to $3$ and large interactions $\eta$ the mass
approaches 3 times the bare mass $m_e$.}

\end